# Positive Exchange Bias between Permalloy and Twined $(10\bar{1}0)$-$Cr_2O_3$ Films

Wei Yuan[1,2†], Tianyu Wang[1,2†], Tang Su[1,2], Qi Song[1,2], Wenyu Xing[1,2], Yangyang Chen[1,2], and Wei Han[1,2*]

[1]International Center for Quantum Materials, Peking University, Beijing, 100871, P. R. China

[2]Collaborative Innovation Center of Quantum Matter, Beijing 100871, P. R. China

[†]These authors equally contribute to this paper.

*Correspondence to be addressed to: weihan@pku.edu.cn (W.H.)

**Abstract:**

We report the discovery of a positive exchange bias between $Ni_{80}Fe_{20}$ (Py) and twined $(10\bar{1}0)$-$Cr_2O_3$ film near its blocking temperature ($T_B$) when it is cooled in an in-plane magnetic field applied along 45 degrees from the two spin configurations of the Cr atoms. This is an abnormal behavior compared to the negative exchange bias at all temperatures below $T_B$ when the cooling and measuring magnetic fields are applied along one of the two spin configurations of the Cr atoms. We speculate these results could be related to the exchange interactions between the twined structure of the $(10\bar{1}0)$-$Cr_2O_3$ film epitaxially grown on the rutile (001)-$TiO_2$ substrate.

## Ⅰ. INTRODUCTION

The interfacial exchange interaction between a ferromagnet (FM) and an antiferromagnet (AFM) gives rise to the exchange bias ($H_{EB}$), the shift of the magnetization hysteresis loop from zero magnetic field [1-4]. There has been enormous effort to explore the exchange bias for practical applications in magnetic sensors and memory devices [5]. In most cases for the FM/AFM bilayer structure, a positive cooling magnetic field gives rise to a negative $H_{EB}$, the shift of the magnetization hysteresis loop toward negative magnetic fields [2]. Whilst, there have been also quite a few studies reporting a positive $H_{EB}$ at the interface of FM/AFM and several mechanisms have been proposed to account for the positive $H_{EB}$ [6-12]. For example, magnetic ordering of the AFM spins under large cooling magnetic fields and the antiferromagnetic coupling between FM/AFM spins could give rise to a positive $H_{EB}$ [6-8]. The net effect of a competition between exchange coupling and uniaxial magnetic anisotropy could lead to a positive $H_{EB}$ [9,10]. Besides, a positive $H_{EB}$ has also been observed in AFM CuMn blow its spin glass transition, which is related to the Ruderman-Kittel-Kasuya-Yosida interaction [11], in (111) oriented $LaNiO_3/LaMnO_3$ superlattice, which is associated with the antiferromagnetic structure in $LaNiO_3$ and the presence of interface asymmetry with $LaMnO_3$ [12], and in the heterostucture of Co/CoO due to interfacial pinning of the AFM spins [13].

In this paper, we report the observation of a positive $H_{EB}$ in the bilayer structure of $Ni_{80}Fe_{20}$ (Py)/$(10\bar{1}0)$-$Cr_2O_3$, where $Cr_2O_3$ is a twined film epitaxailly grown on (001)-$TiO_2$ substrate using pulsed laser deposition. This positive $H_{EB}$ is observed at the temperatures near the blocking temperature ($T_B$) of the $Cr_2O_3$ film and when it is cooled in an in-plane magnetic field directed along the $TiO_2$ [110] direction. On the other hand, if the cooling field is applied along the $TiO_2$ [100] direction, negative $H_{EB}$ is observed for the whole temperature range below $T_B$. This unusual behavior near the blocking temperature indicates complicated spin structures due to the exchange interaction between the twined structure of the $Cr_2O_3$ film and Zeeman interaction associated with an off c-axis magnetic field.

## Ⅱ. EXPERIMENTAL DETAILS

The sample in our experiment is a multilayer structure of Al (20 nm)/Py (10 nm)/ $(10\bar{1}0)$-$Cr_2O_3$ (13 nm)/ (001)-$TiO_2$. The $Cr_2O_3$ is an AFM film with a twined crystalline structure grown on the rutile (001)-$TiO_2$ substrate using high vacuum pulsed laser deposition. The Py and Al layers are grown by magnetron sputtering, where the Al layer is used to protect the Py layer from oxidation during the measurement. Figs.1(a) and 1(b) show the *in-situ* reflection high energy electron diffraction (RHEED) patterns of the rutile $TiO_2$ substrate viewed along [100] and [110] directions at room temperature prior to the $Cr_2O_3$ growth. After the growth of 13 nm $Cr_2O_3$ at 350°C, the RHEED patterns at room temperature are shown in Figs. 1(c) and 1(d) viewed from the same directions as Figs. 1(a) and 1(b), respectively. Fig. 1(e) is the schematic drawing of the side view for the $(10\bar{1}0)$-$Cr_2O_3$ grown on (001)-$TiO_2$ substrate, where the spin orientations of the Cr atoms are aligned in-plane along the $TiO_2$ [100] or [010] directions. The top view of the spin orientations is shown in Fig. 1(f). As there are two equivalent configurations or orientations, with a 50% probability for each, a twined crystalline $Cr_2O_3$ film is resulted. The detailed high-resolution transmission electron microscopy of the interface and this twined crystalline structure has been reported in our previous study [14].

The magnetic hysteresis curves of the multilayer structure are measured via Quantum Design's Magnetic Physical Measurement System (MPMS). Prior to the exchange bias measurement, the sample is first cooled down from 350 K in an in-plane magnetic field of 1000 Oe to set the exchange bias direction. In our study, there are two measurement geometries, as shown in Figs. 2(a) and 2(b). For the first one which we refer to the 1st geometry, the applied cooling and measuring magnetic fields are along the $TiO_2$ [100] direction, which is also one of the two spin configurations of the Cr atoms of $Cr_2O_3$ film (Fig. 2(a)). For the other one which we refer to the 2nd geometry, the applied cooling and measuring magnetic fields are along the $TiO_2$ [110] direction, which is 45 degrees from either one of the two spin configurations of the Cr atoms of $Cr_2O_3$ film (Fig. 2(b)).

## Ⅲ. RESULTS AND DISCUSSION

### A. The magnetization hysteresis measurement

Figs. 2(c) – 2(e) show the typical magnetization hysteresis loops of Py/$Cr_2O_3$ bilayer at 60 K, 50 K, and 30 K respectively, where the black curves indicate the measurement in the 1st geometry (Fig. 2(a)), and the red curves indicate the measurement in the 2nd geometry (Fig. 2(b)). The left ($H_{LC}$) and right ($H_{RC}$) coercive fields can be directly obtained from these curves, and the $H_{EB}$ can be determined from $H_{EB} = (H_{LC}+H_{RC})/2$. The red/black dashed lines indicate the $H_{EB}$ for these magnetization hysteresis loops. At the temperatures close to $T_B$ of $Cr_2O_3$, there is a distinct difference between the two measurement geometries. A normal negative $H_{EB}$ is observed in the 1st geometry (black curves in Figs. 2(c) and 2(d)). Whilst, an interesting positive $H_{EB}$ is observed only in the 2nd measurement geometry (red curves in Figs. 2(c) and 2(d)).

As the temperature decreases further down below 40 K, the magnetization hysteresis loops are almost identical for both measurement geometries. Representative curves measured at 30 K are shown in Fig. 2(e).

## B. Temperature dependence of the coercive fields and $H_{EB}$

Fig. 3(a) shows both $H_{LC}$ (black squares) and $H_{RC}$ (red squares) as a function of the temperature measured in the 1st geometry. As the temperature decreases, the absolute values of both $H_{LC}$ and $H_{RC}$ show a sharp increase below the temperature of ~ 70 K. For the 1st geometry, the absolute value of $H_{RC}$ is smaller than that of $H_{LC}$, which means that the $H_{EB}$ is negative for all the temperatures below $T_B$. For the 2nd geometry, both the absolute values of $H_{LC}$ (black squares) and $H_{RC}$ (blue squares) show a sharp increase below the temperature of ~ 70 K, as shown in Fig. 3 (b). Different from the 1st geometry, a much sharper increase of the $H_{RC}$ is observed in the 2nd geometry, giving rise to a positive $H_{EB}$ close to the $T_B$.

Fig. 3(c) summarizes the $H_{EB}$ as a function of the temperature for both two measurement geometries where the cooling and measuring magnetic fields are along the $TiO_2$ [100] (black squares) and [110] (blue balls) directions. Clearly, a $T_B$ of ~ 70 K is determined for this 13 nm $Cr_2O_3$ film from both measurement geometries. The measured $T_B$ is much lower compared to the value reported on (0001) oriented bulk $Cr_2O_3$ single crystals [15], which could be associated with the trivial finite-size effects [2,16-18] and some non-trivial finite-size effects related to the crystal defects in the $Cr_2O_3$ film [19]. Most interestingly, a positive $H_{EB}$ is observed at 50, and 60 K when the cooling and measuring magnetic fields are along the $TiO_2$ [110] direction (blue balls).

## C. $H_{EB}$ measurement along the $TiO_2$ [010] and $[1\bar{1}0]$ directions

As there are two crystalline zones in $(10\bar{1}0)$ oriented $Cr_2O_3$, we also measure $H_{EB}$ by cooling the magnetic field along the $TiO_2$ [010] and $[1\bar{1}0]$ directions. Fig. 4(a) shows the temperature dependence of $H_{EB}$ by cooling the magnetic field along the $TiO_2$ [100] and [010] directions.

Almost identical exchange biases are observed at each temperature. Fig. 4(b) shows the temperature dependence of $H_{EB}$ by cooling the magnetic field along the $TiO_2$ [110] and $[1\bar{1}0]$ directions. Very similarly, a positive $H_{EB}$ is observed at the temperatures that are close to the $T_B$.

**D. Cooling magnetic field strength effect on $H_{EB}$**

Next, the effect of cooling magnetic field strength on the $H_{EB}$ is studied. Fig. 5(a) and 5(b) show the temperature dependence of $H_{EB}$ under different cooling magnetic field strengths, for the magnetic fields along the $TiO_2$ [100] and [110] directions, respectively. First of all, the magnitude of the cooling magnetic fields has a negligible effect on the $H_{EB}$ at higher temperatures, and a larger cooling magnetic field leads to a lower value of the $H_{EB}$ at lower temperatures, which could be associated with the spin structure at the interface induced by the cooling magnetic field [7]. Second of all, the positive exchange bias is not affected at all by the large magnetic field up to 70000 Oe, indicating the negligible role from the magnetic field strength.

**E. Possible explanation for the positive $H_{EB}$**

Both the temperature and cooling magnetic field strength dependences of the $H_{EB}$ indicate that the positive $H_{EB}$ is related to the twined structure of the $(10\bar{1}0)$-$Cr_2O_3$ film grown on rutile (001)-$TiO_2$ substrate. To study this, we analyze the $H_{EB}$ using numerical calculation based on the coherent rotation model (Stoner-Wohlfarth type) [20], which has been used to explain the positive $H_{EB}$ resulting from the competition between shape anisotropy and the exchange interaction of the heterostructure [9,10]. If we simply neglect the interaction between the spin orientations of the two twined structures in $Cr_2O_3$, the free energy (*f*) can be described as:

$$f = -\mathrm{H}M_s cos\theta - K_{E1}cos(\theta - \theta_{E1}) - K_{E2}cos(\theta - \theta_{E2}) - K_u cos^2(\theta - \theta_u) \qquad (1)$$

where H is the external magnetic field, $M_s$ is the saturation magnetization, $K_u$ is the uniaxial anisotropy, $K_{E1}$ and $K_{E2}$ are the unidirectional anisotropies of the two different twined structures, θ is the angle between H and the FM magnetization, $\theta_{E1}$ and $\theta_{E2}$ are the angles between the H and the unidirectional anisotropic axes of the twined $Cr_2O_3$ film, and $\theta_u$ is the angle between H and the uniaxial axis of FM layer.

Our numerical simulation shows that the $H_{EB}$ is always negative by cooling the magnetic fields along the $TiO_2$ [100] and [010] directions, which are consistent with our experimental results. For the numerical simulation of the $H_{EB}$ by cooling and applying the external magnetic fields along the $TiO_2$ [110] and $[1\bar{1}0]$ directions, we can't reproduce the positive $H_{EB}$ near $T_B$. There could be two mechanisms to account for this behavior. The first one is related to the respective magnetic domains sizes of the FM and AFM layers [21,22]. If the twined regions of AFM are large enough, the Py on top may have different magnetization directions, which would be equivalent to dropping one of the $K_E$ terms in equation (1). As the temperature changes close to $T_B$, the ratio of $K_E/K_u$ could give rise to a positive exchange bias similar to previous report [9]. However, we do not observe a double magnetic hysteresis as shown in ref [21,22], thus, the direct imaging of the magnetic domains would be necessary to test this mechanism. The second mechanism could be relate to the interaction between the two twined structures that is simply neglected in equation (1). This unusual behavior near the blocking temperature indicates complicated spin structures as a result of weak exchange interaction and Zeeman interaction associated with an off c-axis magnetic field. Future theoretical studies are needed to fully understand this behavior.

## Ⅳ. CONCLUSION

We report a positive $H_{EB}$ near the blocking temperature of the $(10\bar{1}0)$-$Cr_2O_3$ film, when the magnetic field is applied along 45 degrees between the two sets of spin configurations of the Cr atoms. This observation indicates complicated spin structures at the interface between Py and twined $(10\bar{1}0)$-$Cr_2O_3$ film near its $T_B$.

**Acknowledgements**

We acknowledge the fruitful discussion with Dr. Axel Hoffmann and Prof. Jing Shi, and the funding support of National Basic Research Programs of China (973 Grants 2014CB920902, 2015CB921104, and 2013CB921903) and the National Natural Science Foundation of China (NSFC Grant 11574006). Wei Han also acknowledges the support by the 1000 Talents Program for Young Scientists of China.

**Figure Captions**

Figure 1: (a-b) RHEED patterns of rutile $TiO_2$ substrate viewed along the [100] and [010] directions, respectively. (c-d) RHEED patterns of the twined $(10\bar{1}0)$-$Cr_2O_3$ film (thickness: 13 nm) viewed from $TiO_2$ [100] and [110] directions. (e) The schematic drawing showing the side view of the atomic interface between $Cr_2O_3$ and $TiO_2$. The red dashed line indicates the interface between $Cr_2O_3$ and $TiO_2$, and the black dashed line indicates the crystalline boundary of the twined $Cr_2O_3$ film. (f) The schematic drawing showing the top view of the twined $Cr_2O_3$ film. The spins of the Cr atoms are indicated by the red arrows.

Figure 2: (a-b) Schematic drawings of the in-plane exchange bias measurement between Py and the twined $Cr_2O_3$ with the magnetic field applied along $TiO_2$ [100] (a; 1st geometry) and [110] (b; 2nd geometry) directions. (c-e) The magnetization curves of Py measured as a function of the in-plane magnetic field at 60K, 50K, and 30K, respectively. The red/black curves indicate the measurement with the magnetic fields applied along $TiO_2$ [110]/[100] direction. The red/black dashed lines are the guide lines for $H_{EB}$.

Figure 3: (a-b) Left ($H_{LC}$) and right ($H_{RC}$) coercive fields of Py as a function of the temperature for the magnetic fields applied along $TiO_2$ [100] and [110] directions, respectively. (c) Temperature dependence of $H_{EB}$ between Py and $Cr_2O_3$ for the magnetic fields applied along $TiO_2$ [100] and [110] directions, respectively. The dashed magenta line is the guide line for zero $H_{EB}$.

Figure 4: (a) Temperature dependence of $H_{EB}$ between Py and $Cr_2O_3$ for the magnetic fields applied along $TiO_2$ [100] and [010] directions, respectively. (b) Temperature dependence of $H_{EB}$

between Py and $Cr_2O_3$ for the magnetic fields applied along $TiO_2$ [110] and $[1\bar{1}0]$ directions, respectively. The dashed magenta lines are the guide lined for zero $H_{EB}$.

Figure 5: (a) Temperature dependence of $H_{EB}$ between Py and $Cr_2O_3$ under cooling magnetic fields of 1000 (black) and 70000 Oe (red) for the magnetic fields applied along $TiO_2$ [100] direction. (b) Temperature dependence of the $H_{EB}$ between Py and $Cr_2O_3$ under cooling magnetic fields of 1000 (black), 30000 (green), and 70000 Oe (blue), respectively, for the magnetic fields applied along $TiO_2$ [110] direction. The dashed magenta lines are the guide lined for zero $H_{EB}$.

Figure 1

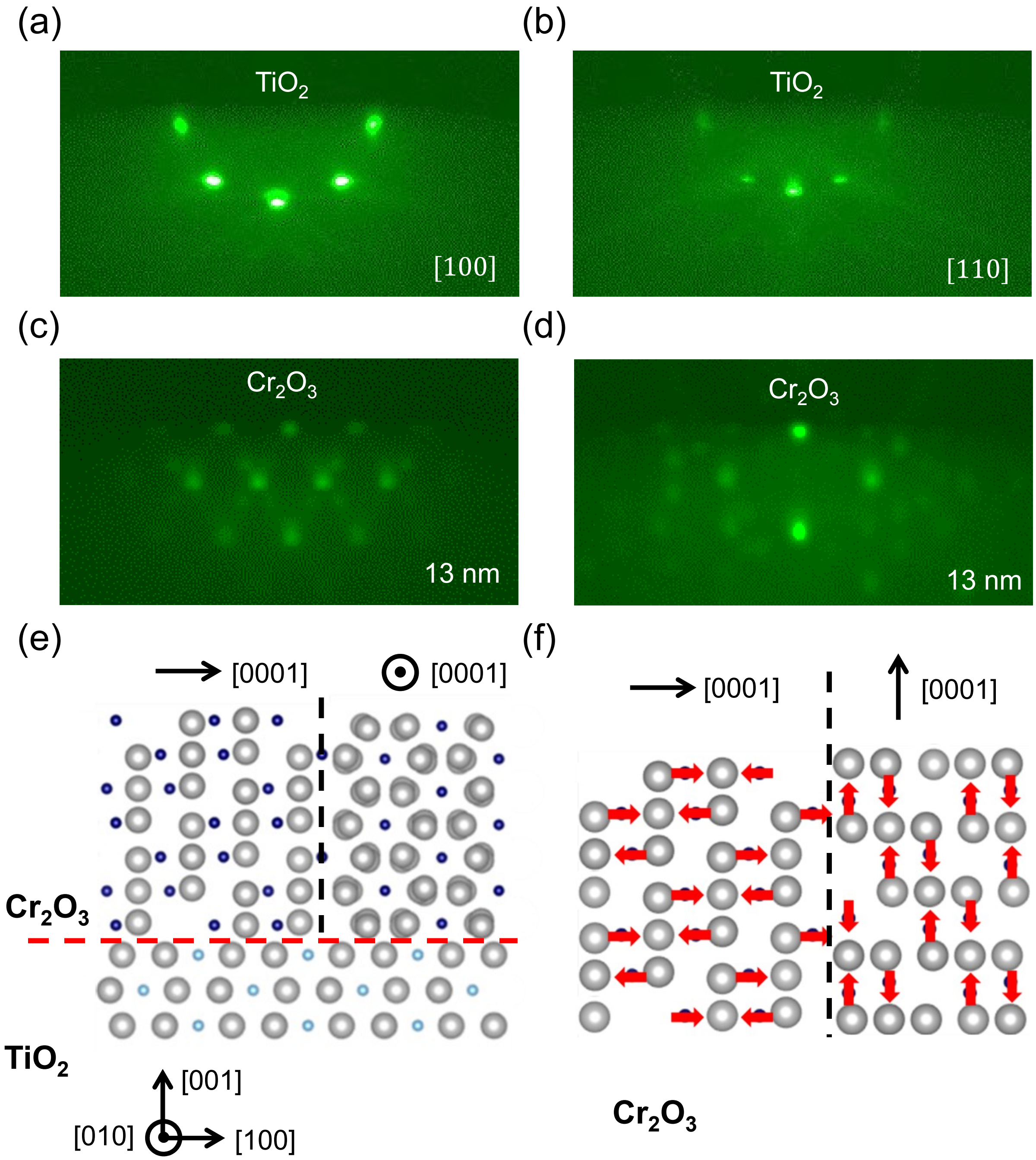
(a)
$TiO_2$
[100]
(b)
$TiO_2$
[110]
(c)
$Cr_2O_3$
13 nm
(d)
$Cr_2O_3$
13 nm
(e)
[0001]
[0001]
$Cr_2O_3$
$TiO_2$
[001]
[010]
[100]
(f)
[0001]
[0001]
$Cr_2O_3$

Figure 2

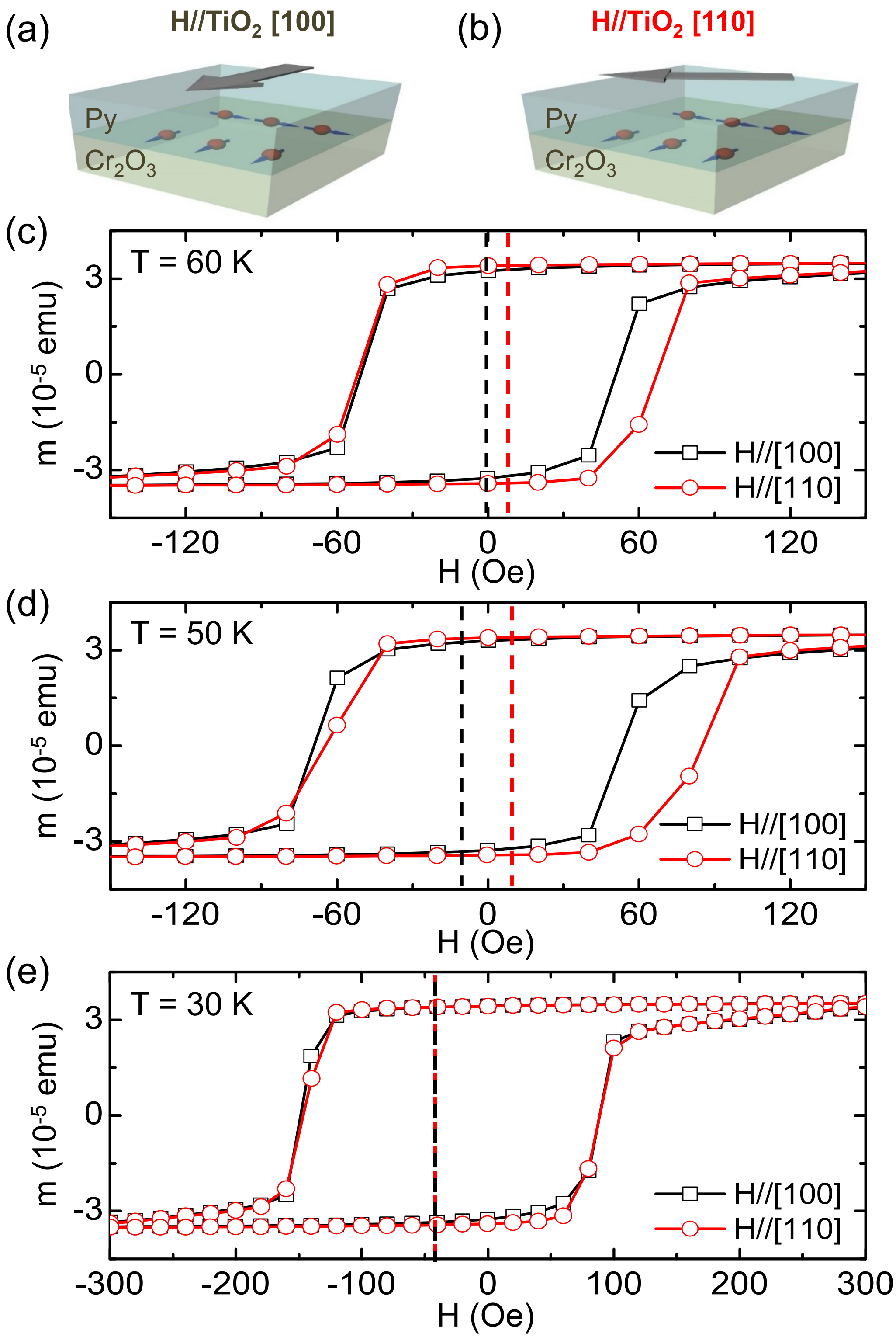

(a)
H//TiO2 [100]
Py
Cr2O3
(b)
H//TiO2 [110]
Py
Cr2O3
(c)
T = 60 K
m (10-5 emu)
H (Oe)
H//[100]
H//[110]
(d)
T = 50 K
m (10-5 emu)
H (Oe)
H//[100]
H//[110]
(e)
T = 30 K
m (10-5 emu)
H (Oe)
H//[100]
H//[110]

Figure 3

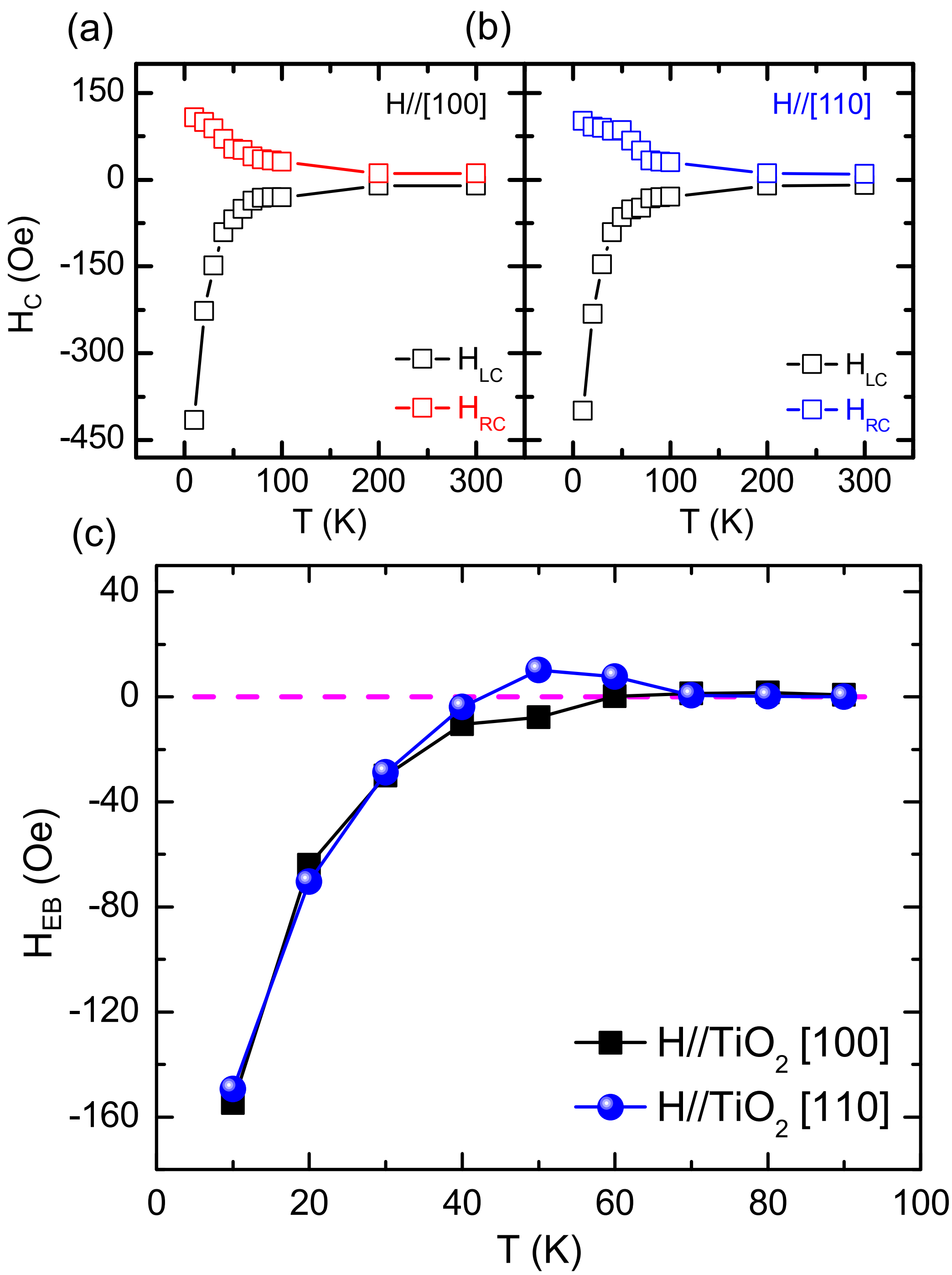

(a)
(b)
H//[100]
H//[110]
150
0
-150
-300
-450
$H_C$ (Oe)
$H_{LC}$
$H_{RC}$
0
100
200
300
T (K)
(c)
40
0
-40
-80
-120
-160
$H_{EB}$ (Oe)
H//TiO$_2$ [100]
H//TiO$_2$ [110]
0
20
40
60
80
100
T (K)

Figure 4

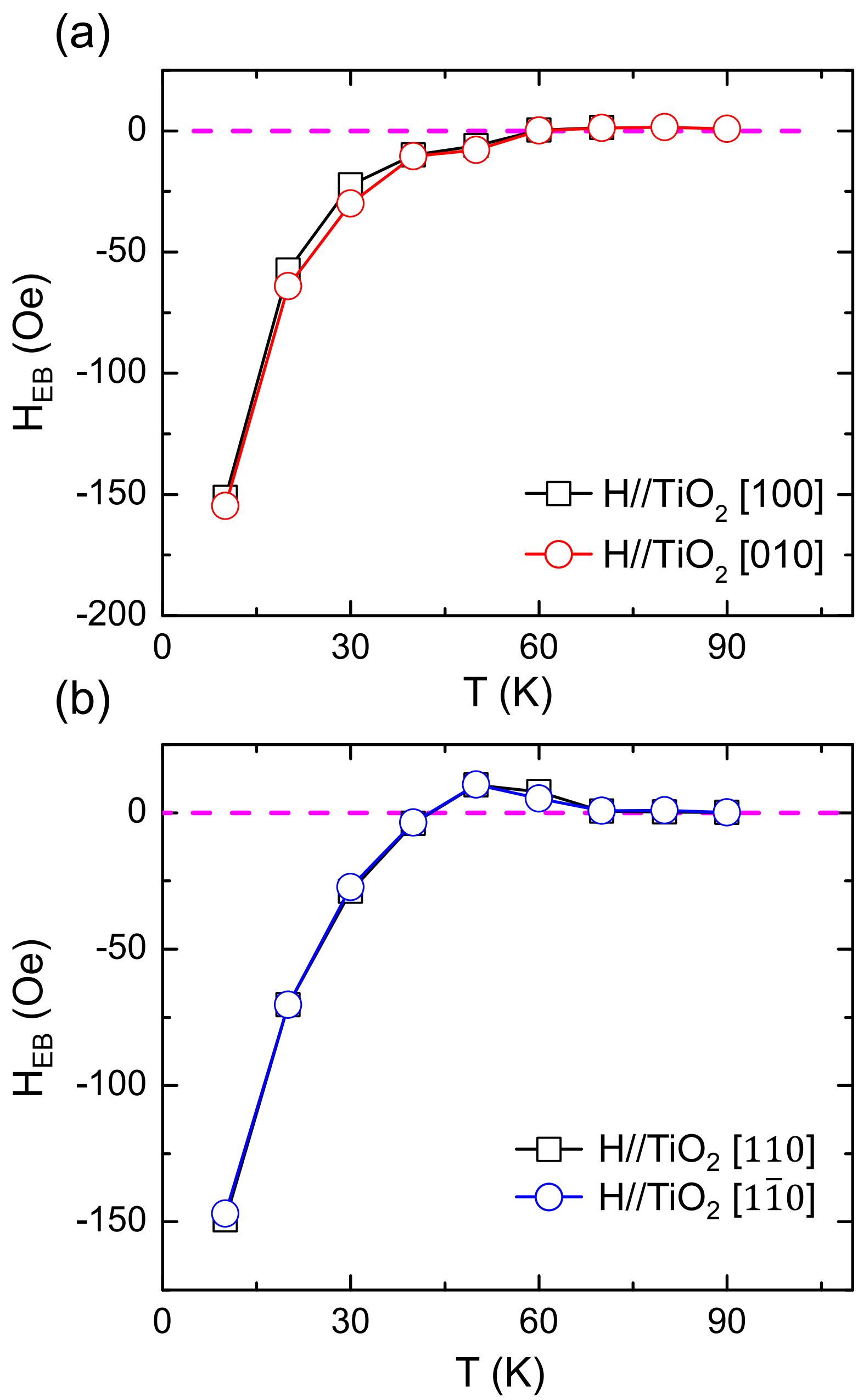

(a)
0
-50
-100
-150
-200
$H_{EB}$ (Oe)
0
30
60
90
T (K)
H//$TiO_2$ [100]
H//$TiO_2$ [010]
(b)
0
-50
-100
-150
$H_{EB}$ (Oe)
0
30
60
90
T (K)
H//$TiO_2$ [110]
H//$TiO_2$ [1$\bar{1}$0]

Figure 5

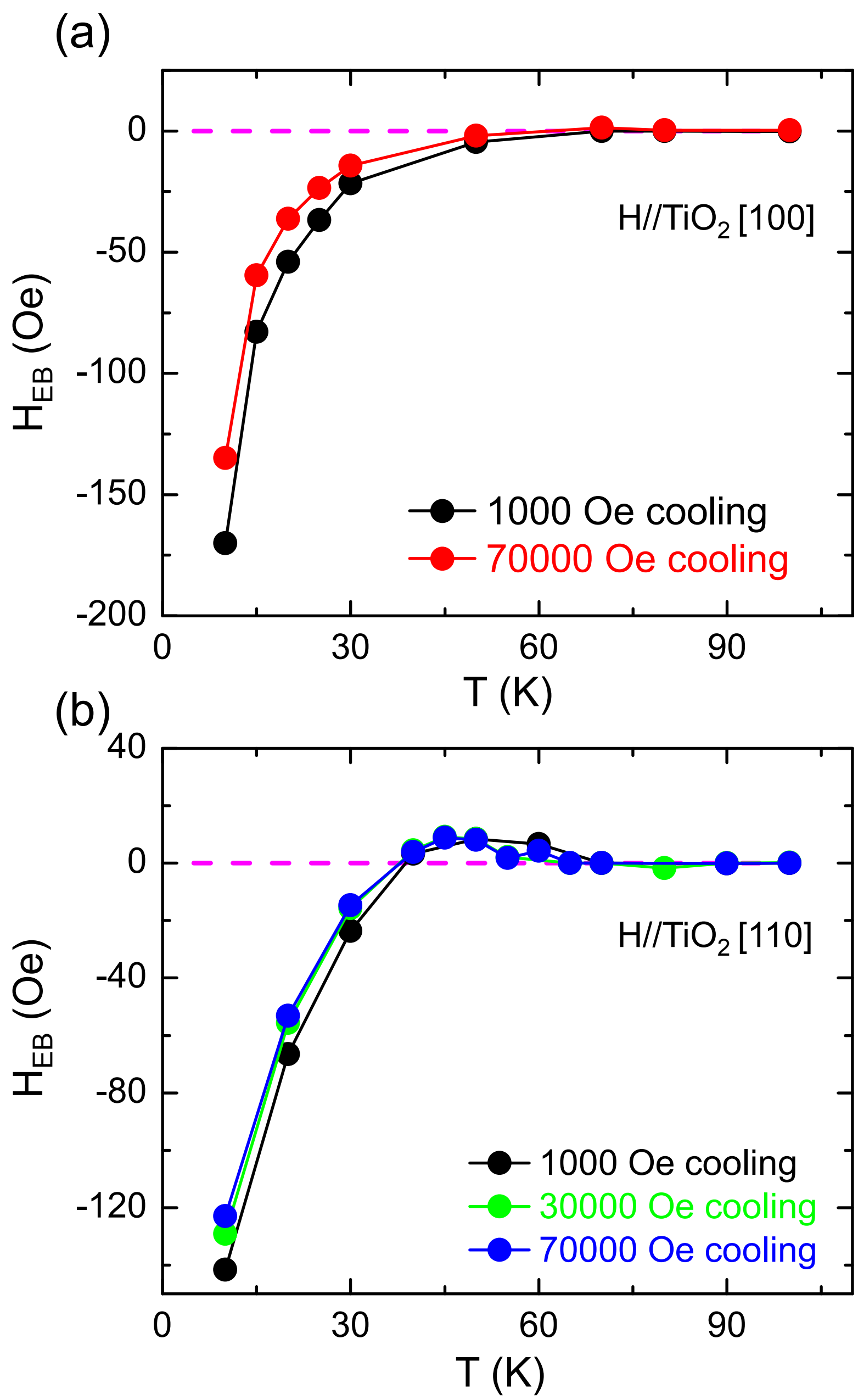

(a)
H//TiO2 [100]
1000 Oe cooling
70000 Oe cooling
H_EB (Oe)
T (K)
(b)
H//TiO2 [110]
1000 Oe cooling
30000 Oe cooling
70000 Oe cooling
H_EB (Oe)
T (K)